\documentclass[11pt]{article}
\usepackage{iap2000,epsfig}

\def\be{\begin{equation}}
\def\ee{\end{equation}}
\def\bea{\begin{eqnarray}}
\def\eea{\end{eqnarray}}

\begin{document}
\vspace*{4cm}
\title{The Luminosity Function of 81 Abell Clusters from the CRoNaRio catalogues}

\author{S. PIRANOMONTE}
\address{Osservatorio Astronomico di Capodimonte,
via Moiariello 16, 80131 Napoli, Italy}
\author{M. PAOLILLO} 
\address{D.S.F.A., Universit\'a di Palermo, Via Archirafi 36, 90123 Palermo}
\author{S. ANDREON, G. LONGO, E. PUDDU}
\address{Osservatorio Astronomico di Capodimonte,
via Moiariello 16, 80131 Napoli, Italy}
\author{R. SCARAMELLA}
\address{Osservatorio Astronomico di Roma, Via Frascati 33, 00040 Roma} 
\author{R. GAL, S.G. DJORGOVSKI }
\address{Department of Astronomy, Caltech, USA}

\maketitle

\abstracts{We present the composite luminosity function (hereafter LF) of galaxies for 81 Abell clusters studied in our survey of the Northern Hemisphere, using DPOSS data processed by the CRoNaRio collaboration. The derived LF is very accurate due to the use of homogeneous data both for the clusters and the control fields and to the local estimate of the background, which takes into account the presence of large-scale structures and of foreground clusters and groups.
The global composite LF is quite flat down to $M^*+5$ has a slope 
$\alpha\sim-1.0\pm0.2$ with minor variations from blue to red filters, and $M^*\sim-21.8,-22.0,-22.3$ mag ($H_0=50$ km s$^{-1}$ Mpc$^{-1}$) in the $g, r$ and $i$
filters, respectively (errors are detailed in the text).
We find a significant difference between rich and poor clusters thus arguing in favour of a dependence of the LF on the properties of the environment.
}
\begin{figure*}[]
\resizebox{\hsize}{!}{\includegraphics{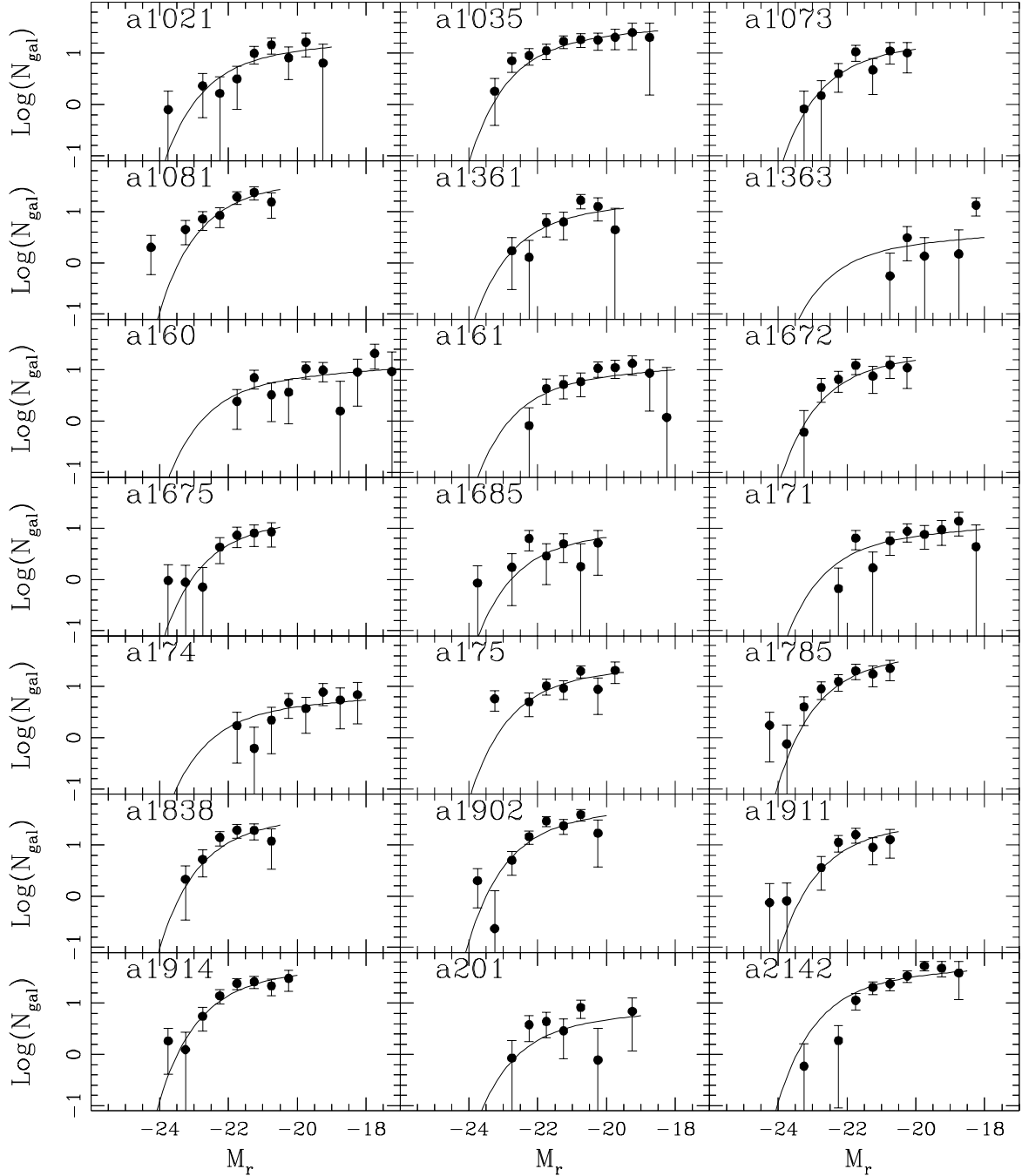}}

\caption[]{{\bf (a)} 
The background-corrected galaxy counts for each cluster of our sample in the $r$ band (i.e. the LF). The best-fit Schechter function of the cumulative LF (par.3), normalized to the total counts in each cluster, is shown as a continous line.}
\end{figure*}
\begin{figure*}[]
\resizebox{\hsize}{!}{\includegraphics{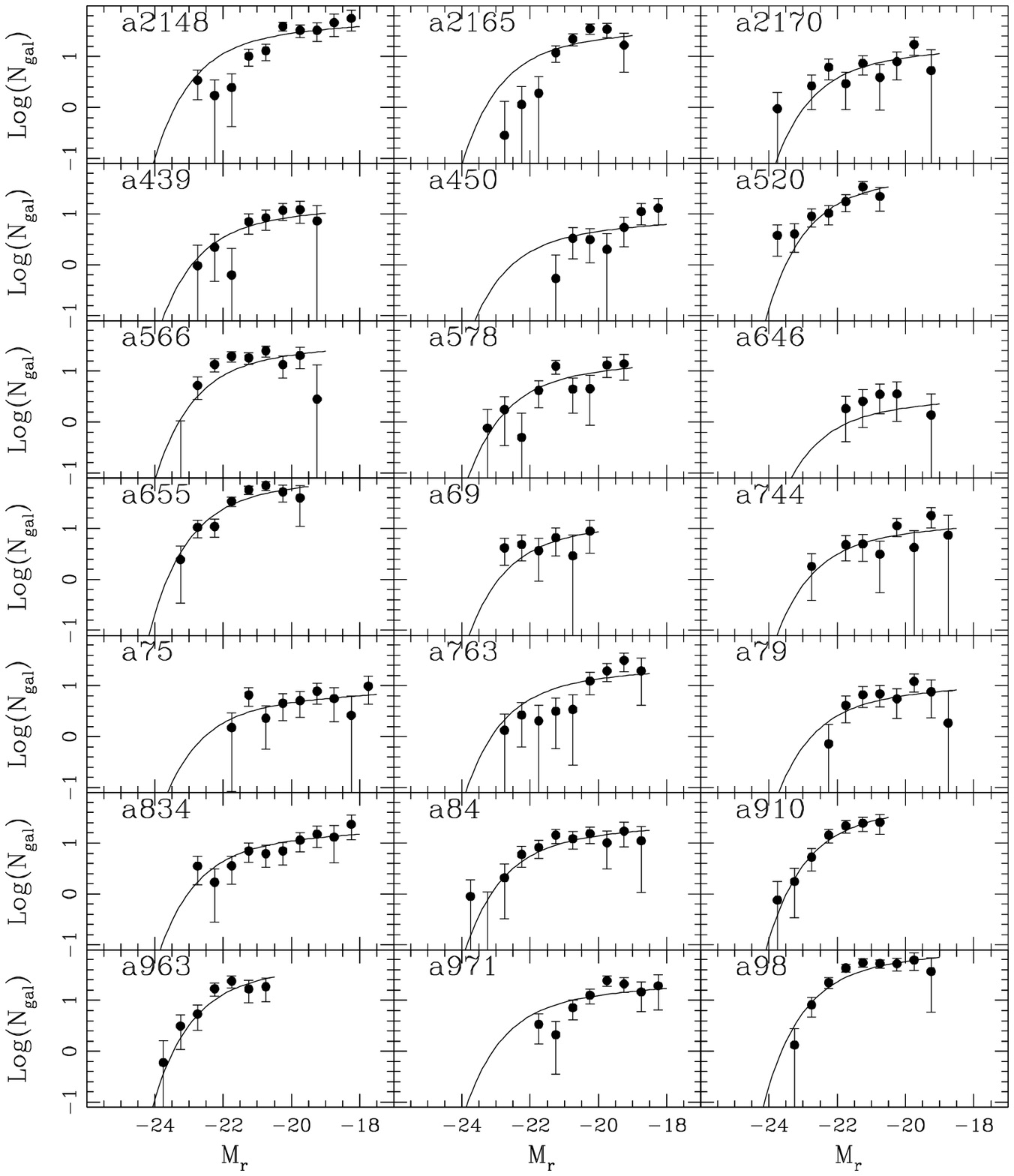}}
\addtocounter{figure}{-1}
\caption[]{{\bf (b)}
Fig. (a) cont.}
\end{figure*}



\begin{figure}[th]
\begin{center}
\epsfysize=14cm 
\epsfxsize=14cm 
\hspace{0.cm}\epsfbox{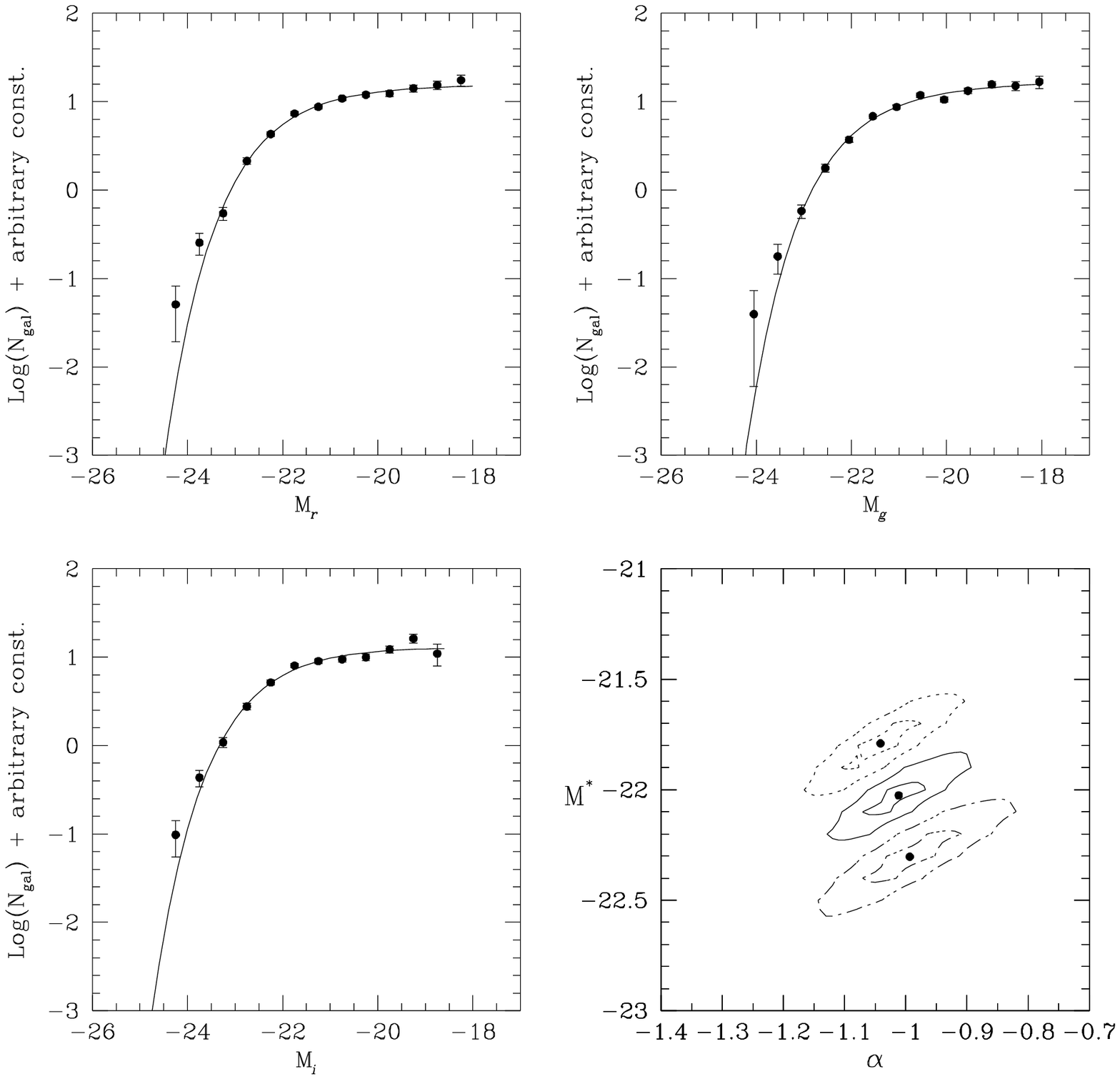} 
\end{center}
\caption[h]{\it{LF for 81 clusters obtained excluding the brightest member of each cluster (filled dots). The best fit Schechter functions are represented by the continuos line, with the 68\% and 99\% confidence levels of the best fit parameters in the bottom right panel ($g$:dotted line; $r$ continuos line; $i$: dashed-dotted line)}.}
\end{figure}

\section{Introduction}
The galaxy luminosity function measures the relative frequency of galaxies as a function of luminosity.
Cluster LF's can be determined as the statistical excess of galaxies along the line of sight of the clusters with respect to control field directions due to the fact that clusters appear as overdensities with respect to the field. This approach assumes that the contribution of the fore-background galaxies along the line of sight of the clusters is equal to an ``average'' value. An hypotesis that is rather weak since very often nearby groups, clusters or superclusters happen to lay in same the direction and therefore affect the determination of the LF. 

This  problem is even more relevant when sampling the cluster outskirts due to the following problems: i) the low galaxy density of the outer regions is strongly affected even by small contaminations; ii) the large observing area makes more probable the presence of contaminating structures. It needs to be stressed that these outer regions are very relevant due to the fact that they are the putative places for galaxy evolution to occur (van Dokkum et al. 1998).

The precence of non zero correlation function makes very time consuming the accurate determination of LF's using traditional CCD imagers (due to their small field of view) and obliges most authors to use comparison average field counts taken from the literature. These average counts are usually obtained from regions of the sky completely unrelated to the cluster position. Alternative routes are either to observe small comparison fields or to recognise individual cluster membership either on spectroscopic or colour grounds.

Wide field (hereafter WF) imagers such as Schmidt plates or large format CCD's are therefore the ideal tools to perform accurate determinations of LF's for statistically significant samples of clusters. In this paper we present intermediate results from a long term project aimed to derive LF's for a large and statistically well defined sample of Abell clusters.

The work is done in the framework of the CRoNaRio collaboration (Weir et al. 1995, Andreon 1997, Djorgovsky et al. 1998) aimed to produce the first complete catalogue of all object visible on the Digitised Palomar Sky Survey (hereafter DPOSS). In what follow we discuss the results for a subsample of 81 clusters.

\section{Individual LF determination}

The LF is computed statistically as the difference between galaxy counts in cluster direction and those in a ``control field'' direction.

Therefore, in order to compute the cluster LF, the first step is to subtract the background contamination. The method, together with the errors involved in the subtraction process, is detailed in Paolillo et al. (1999) and in Paolillo et al. (2000). Field counts are measured around each cluster, thanks to the wide field coverage of DPOSS plates, after removing density peaks. A ``local field'', measured all around the studied clusters, is the adopted estimate of the
background counts in the cluster direction.  This procedure allows a better measure of the contribution of
background galaxies to counts in the cluster direction than the usual single control field,
since it allows to correct for the presence of possible underlying large-scale structures
both at the cluster distance and in front or behind it.

Once we have searched for the central $1.5\sigma$ density peak in the central region (cf. Paolillo et al. 2000) we then derive the cluster LF by subtracting from the galaxy counts the local field counts, rescaled to the effective cluster area. This approach allows to take into account the cluster morphology without having to adopt a fixed cluster radius, and thus to apply the local field correction to the region where the signal (cluster) to noise (field and cluster fluctuations) ratio is higher, in order to minimize statistical uncertainties.

Fig.1a,b show the LFs for 42 individual clusters in the $r$ band. The other 39 LFs are presented in Paolillo et al. (2000). The solid line is the Schechter function with $\alpha$ =  -1.03 $^{+0.09}_{-0.07}$, $M^* =-22.03\pm0.16$ and the parameter $\phi^*$ normalized to the total counts in each cluster.

   \begin{table}
   \caption[]{LF best fit Schechter parameters. The given errors are
		referred to 1$\sigma$ confidence levels.}   
\begin {center}   
\begin{tabular}{cccc}
\hline
\\
{Band} & {$\alpha$} & {$M^*$} & {$\chi^2$/d.o.f.}\\
\\
\hline
\\

$g$ & -0.99$^{+0.09}_{-0.07}$ & -21.82$^{+0.13}_{-0.17}$ & 9.1/13\\ 
$r$ & -1.03$^{+0.09}_{-0.07}$ & -22.03$\pm$0.16 & 10.9/13 \\
$i$ & -0.99$^{+0.12}_{-0.11}$ & -22.30$\pm$0.20 & 11.1/12\\

\hline
\end{tabular}
\end {center}
\end{table}

\section{Results}

We combine individual LF's of many clusters to obtain a composite LF (CLF) for all clusters in
our sample. We adopted the method in  Garilli et al. (1999).  In practice, the composite LF
is obtained by weighting each cluster by its  richness which is defined as the number of galaxies brighter than -17.25 computed in an opportune isoplate. The preliminar CLF for a first sample of 81 clusters is shown in Figure 2.

The fit of the CLF to a Schechter (1976) function gives the values in Table 1 where $M^*$ is the characteristic knee magnitude and $\alpha$ is the slope of the LF at faint magnitudes. In our magnitude range the CFL turns out to be quite well represented by a Schechter function (see $\chi^2$ in Tab.1).
  
Figure 2 shows the three best-fit functions and the $68\%$ and $90\%$ confidence levels. The best fit parameters of the three filters differ by more than 90\% confidence level (c.l.), mainly because $M^*$ become brighter with wavelenght.
It needs to be taken into account that the derived colour term is indicative only, since we used isophotal-corrected magnitudes which -due to the different filters- are not homogeneous in the different filters.  

\begin{figure*}[]
\begin{center}
\epsfysize=7cm 
\epsfxsize=14cm 
\hspace{0.cm}\epsfbox{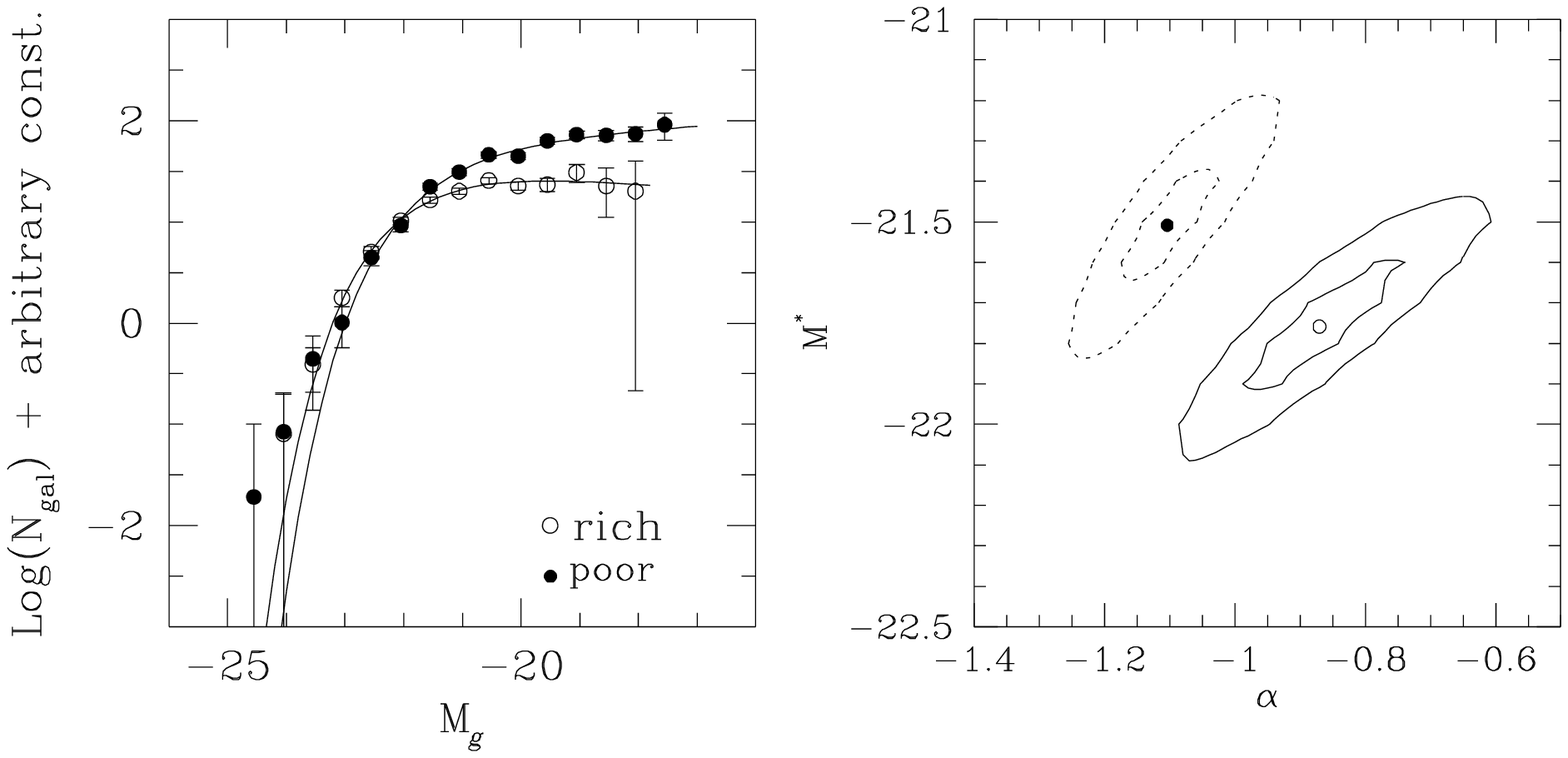} 
\end{center}
\caption[h]{\it{Left panel: the LF and the best fit Schechter functions of the rich (R $>$1) and poor ($R \leq 1$) Abell clusters in the $g$ band. Right panel: 68\% and 99\% confidence levels relative to the
fit parameters ($R > 1$: continous lines; $R \leq 1$: dotted lines). The normalization is arbitrary.}}
\vskip 1cm
\begin{center}
\epsfysize=7cm 
\epsfxsize=14cm 
\hspace{0.cm}\epsfbox{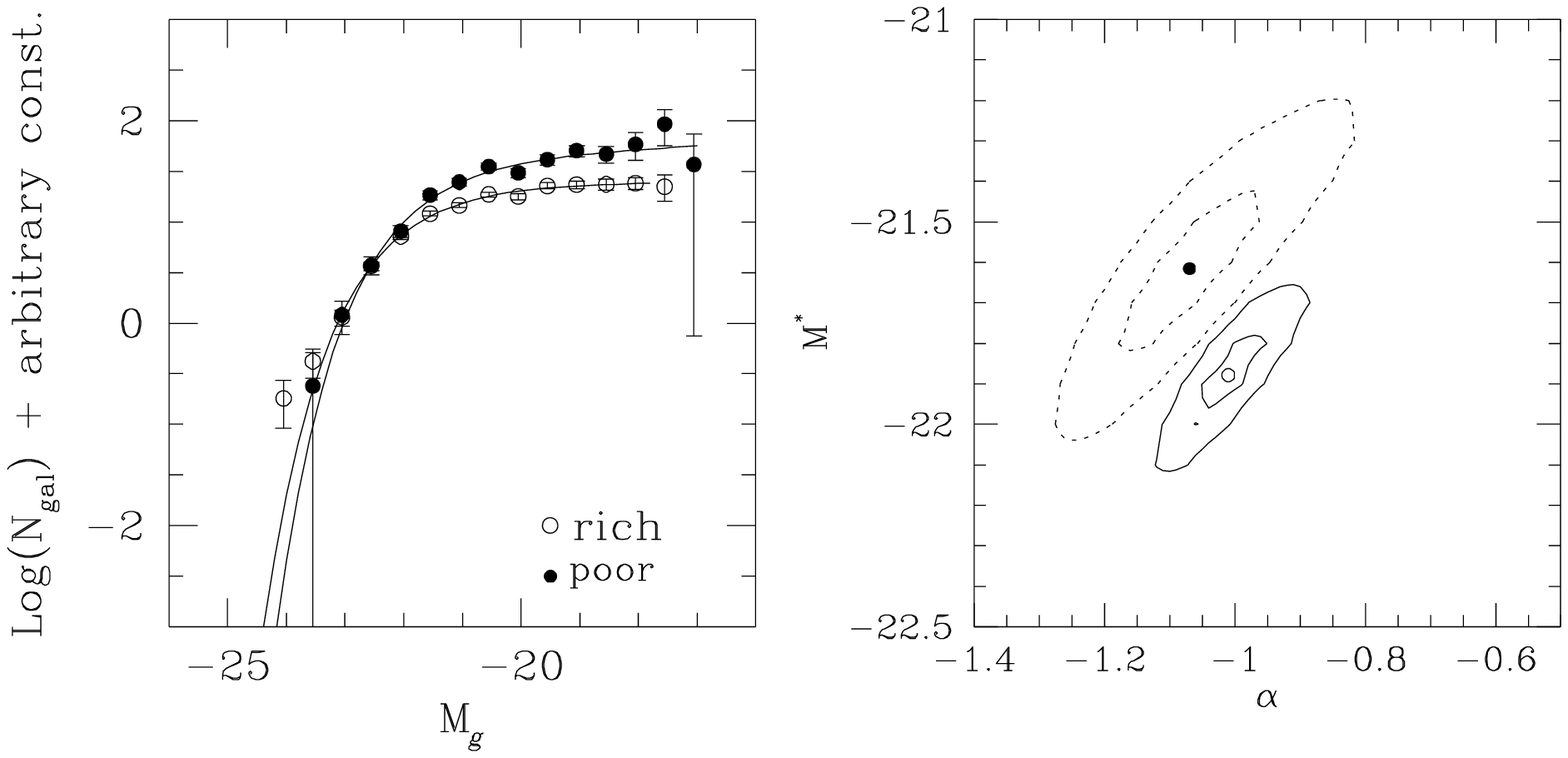} 
\end{center}
\caption[h]{\it{Left panel: the LF and the best fit Schechter functions of rich and poor clusters in the $g$ band, where the richness is directly determinated by us. Right panel: 68\% and 99\% confidence levels relative to the fit parameters (rich: continous lines; poor: dotted lines). The normalization is arbitrary.}}
\end{figure*}

\subsection{Dependence on the environment}

The relatively large number of clusters used in the present study allowed us to
investigate the dependence of the LF on the cluster richness parameter. 

We first divided our sample in two richness subclasses (R$>$1 and R$\leq1$) where R is defined as in Abell (1958).

Fig. 3 shows that Abell rich clusters (R$>$1)  have flatter slope of the LF, at $3\sigma$ c.l., than poor ones (R$\leq1$). This confirms that poor clusters host a larger fraction of dwarfs as suggested by Driver, Couch, \& Phillips (1988), Andreon (2000) and Garilli et al.(1999). 

When instead the sample is divided according to our richness parameter (Fig. 4)which is defined as the number of galaxies brighter than -17.25 computed in an opportune isoplate, we still find a statistical significance difference but the main difference is in $M^*$: rich clusters turn out to have brighter $M^*$. Whether this effect is real or it is just a selection effect, is still under investigation.

\section{Conclusions}
We computed the composite LF of 81 clusters of galaxies at $0.08<z<0.4$ in three filters from the DPOSS plates, by using the fact that clusters are galaxy overdensities with respect to the field. The LF is well described by a Schechter function, with a shallow slope $\alpha \sim -1.0$ with minor variation from blu to red filters and $M^* \sim -21.8, -22.0, -22.3$ in $g$, $r$ and $i$ filters respectively. The best-fit values of $M^*$ increases from the blue to the red band as is expected from the color of the dominant population in clusters.

When the clusters sample is divided in richness systematic differences are found.

\end{document}